%% file: P-doped_Ba122_PRL.tex
\documentclass[aps,twocolumn,superscriptaddress,showpacs,amssymb,prl,floatfix,preprintnumbers]{revtex4-1}
\usepackage{graphicx,color}
\usepackage{epstopdf}
\usepackage[usenames,dvipsnames,svgnames,table]{xcolor}

\usepackage{hyperref}

\hypersetup{colorlinks=true,linkcolor=blue,citecolor=blue,urlcolor=blue}

\newcommand{\slrr}{$T_1^{-1}$}

\newcommand{\bafeasp}{BaFe$_2$(As$_{1-x}$P$_x$)$_2$}

\begin{document}

\thispagestyle{myheadings}

\title{NMR evidence for inhomogeneous nematic fluctuations in BaFe$_2$(As$_{1-x}$P$_x$)$_2$}

\author{A. P. Dioguardi}
\author{T. Kissikov}
\author{C. H. Lin}
\author{K. R. Shirer}
\author{M. M. Lawson}

\affiliation{Department of Physics, University of California, Davis, CA 95616, USA}
\author{H.-J. Grafe}
\affiliation{IFW Dresden, Institute for Solid State Research, P.O. Box 270116, D-01171 Dresden, Germany}
\author{J.-H. Chu}
\author{I. R. Fisher}
\affiliation{Department of Applied Physics and Geballe Laboratory for Advanced Materials, Stanford University, Stanford, California 94305, USA}
\affiliation{Stanford Institute of Energy and Materials Science, SLAC National Accelerator Laboratory, 2575 Sand Hill Road, Menlo Park, California 94025, USA}
\author{R. M. Fernandes}
\affiliation{School of Physics and Astronomy, University of Minnesota, Minneapolis, Minnesota 55116, USA}
\author{N. J. Curro}
\affiliation{Department of Physics, University of California, Davis, CA 95616, USA}

\date{\today}

\begin{abstract}

We present evidence for nuclear spin-lattice relaxation driven by glassy nematic fluctuations in isovalent P-doped BaFe$_2$As$_2$ single crystals.  Both the $^{75}$As and $^{31}$P sites exhibit stretched-exponential relaxation similar to the electron-doped systems. By comparing the hyperfine fields and the relaxation rates at these sites we find that the As relaxation cannot be explained solely in terms of magnetic spin fluctuations.  We demonstrate that nematic fluctuations couple to the As nuclear quadrupolar moment and can explain the excess relaxation. These results suggest that glassy nematic dynamics are a universal phenomenon in the iron-based superconductors.
\end{abstract}

\maketitle

The iron-based superconductors continue to attract broad interest
not only because of the presence of unconventional high-temperature
superconductivity, but also because of their unusual normal state
behavior \cite{Johnston:2010cs}. As in other unconventional superconductors
the superconductivity emerges at the boundary of antiferromagnetism,
suggesting an important role for antiferromagnetic fluctuations in
the superconducting pairing mechanism \cite{uedareview,chubukovreview}.
In recent years, however, significant attention has focused on the
presence of nematic order that breaks the $C_{4}$ point group symmetry
of the lattice {at} the tetragonal to orthorhombic
structural transition, as well as the large nematic susceptibility
in the {tetragonal} paramagnetic regime \cite{Fang:2008by,IsingSpinOrderSachdev2008,Fernandes:2014en,Gallais:2013fj,Chu:2010hg}.
Elastoresistance measurements indicate that the static nematic susceptibility diverges
near optimal doping in several pnictide families \cite{Chu:2012ez,Kuo2015}.
Similar conclusions are drawn from both elastic constant measurements \cite{Bohmer:2014jv} and Raman spectroscopy \cite{Kretzschmar2015,Gallais2015,Thorsmoelle2014}.
An open question, therefore, is whether there is a connection between
the nematic fluctuations and the unconventional superconductivity
in these materials \cite{KivelsonNematicQCP2015,ScalapinoNematicQCP2015,Fernandes:2014en}.

Experimentally the nematic fluctuations appear to be strongly coupled
to the spin degrees of freedom. The shear modulus and the nuclear
spin-lattice relaxation rate, $T_{1}^{-1}$, are strongly temperature
and doping dependent, but scale with one another {in
Co- and K-doped BaFe$_{2}$As$_{2}$ }\cite{Fernandes:2013ib,bohmer2015electronic}.
Assuming that the dominant channel for \slrr\ is via the hyperfine
coupling to the Fe spins, this empirical relationship implies that
the lattice and spin fluctuations have a common origin. Further evidence
for a coupling between these order parameters has emerged from neutron
scattering studies which reveal that $C_{4}$ symmetry is broken for
the spin fluctuations in the {high-temperature} phase
of uniaxial-strained {Ba(Fe$_{1-x}$Ni$_{x}$)$_{2}$As$_{2}$}
\cite{StrainedPnictidesNS2014science}. Other neutron
scattering experiments have uncovered an enhancement of spin fluctuations
in both LaFeAsO and {Ba(Fe$_{1-x}$Co$_{x}$)$_{2}$As$_{2}$} between
the structural transition and antiferromagnetic transition temperatures
\cite{GoldmanSpinNematicOrderPRL2015}. In contrast, the iron chalcogenide
FeSe undergoes a nematic phase transition {despite
the absence of long-range magnetic order down to the lowest temperatures.
Although nuclear magnetic resonance (NMR) measurements do not observe significant low-energy magnetic
fluctuations above the nematic transition \cite{Baek2015,IshidaFeSePRL2015},
neutron scattering indicates the presence of sizable spin fluctuations
at moderate energies \cite{Boothroyd_FeSe,Wang_FeSe}.}

Direct evidence for the nematic fluctuations has remained elusive. NMR studies of {Ba(Fe$_{1-x}$M$_x$)$_2$As$_2$} (M = Co, Cu) uncovered the presence of glassy spin dynamics extending up to 100 K, with a doping and temperature response that matches that of the nematic susceptibility \cite{Dioguardi:2013dd,Dioguardi:2015tx}. The glassy behavior possibly originates from quenched disorder, which can act as a random local field on the fluctuating nematic order \cite{Kuo2015,RFIMsimulations2010}. In such a case, magnetoelastic coupling ensures that random variations in the local value of the nematic order parameter also affects the local spin fluctuations measured by NMR \cite{Fernandes:2013ib}.

\begin{figure}
\includegraphics[width =\linewidth]{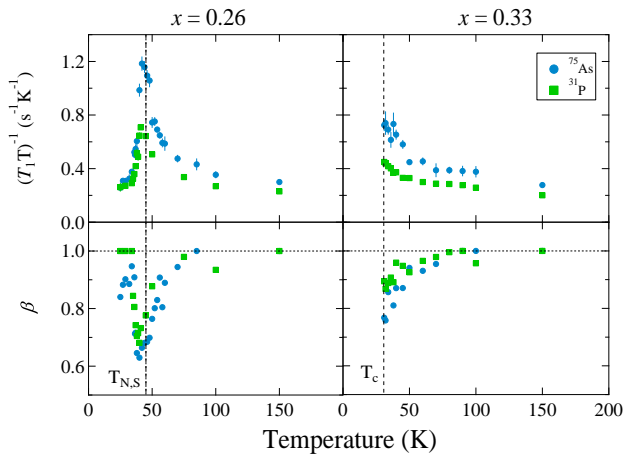}
	\caption{$(T_1T)^{-1}$ and $\beta$ for $^{75}$As ($\bullet$) and $^{31}$P ($\square$) vs. temperature for  $x=0.26$ (underdoped) and  $x=0.33$ (optimal doping). }
	\label{fig:T1Tinv_layout}
\end{figure}

In order to investigate the presence of nematic fluctuations directly,
we have investigated the NMR properties of both the $^{75}$As ($^{75}\gamma=7.2919$
MHz/T, $I={3}/{2}$, $Q=0.31$ barns) and $^{31}$P ($^{31}\gamma=17.2356$
MHz/T, $I={1}/{2}$) in single crystals of \bafeasp. Isovalent substitution
of P for As in BaFe$_{2}$As$_{2}$ gives rise to superconductivity
with a phase diagram that is similar to that of the electron- or hole-doped
system \cite{Nakai:2010ku,Nakai:2013by}. These
nuclei present a unique opportunity because they are located at the
same crystallographic site, but the {{} $I=3/2$ $^{75}$As
nucleus} experiences a quadrupolar interaction whereas the{{}
$I=1/2$ $^{31}$P nucleus} does not. Both nuclei are sensitive to
magnetic hyperfine fluctuations of the Fe spins, however the As is
also sensitive to fluctuations of the local electric field gradient
(EFG). Nematic fluctuations {couple directly to} EFG
and give rise to an extra channel for relaxation at the As. We carefully
analyze the relaxation of both nuclei, and conclude that quadrupolar
fluctuations are indeed contributing to the relaxation of the As,
{giving rise to a maximum in the ratio $^{75}T_{1}^{-1}/^{31}T_{1}^{-1}$
at} the structural transition temperature, $T_{s}$. We also observe
inhomogeneous dynamics that result in stretched exponential spin-lattice
relaxation for both nuclear species. The amount of dynamical inhomogeneity
is similar to previous NMR observations in both Co- and Cu-doped BaFe$_{2}$As$_{2}$
\cite{Dioguardi:2013dd,Dioguardi:2015tx} and LaFeAsO \cite{hammerath:2010ki,hammerath:2013jj}
compounds.

Single crystals were synthesized via a self-flux method and characterized via transport measurements to determine P-doping levels. The P concentration $x$ was estimated by comparison of transport properties with samples from the same and similar growth batches for which the composition had been determined via microprobe analysis \cite{KuoPBa122PRB2012}. The spin-lattice relaxation rates of $^{75}$As and $^{31}$P were measured at the central  transition ($I_z = \pm{1}/{2}$) in two  BaFe$_2$(As$_{1-x}$P$_x$)$_2$ crystals with $x=0.26$ (underdoped, $T_s = T_N = 45$ K) and $x=0.33$  (optimally doped, $T_c = 31$ K) as a function of temperature via a standard inversion recovery pulse sequence. The crystals were aligned with the external field $\mathbf{H}_0 = 11.7285$ T oriented perpendicular to the $c$-axis, and the magnetization recovery was fit to the appropriate normal modes recovery function modified by a stretching exponent $\beta$, as described in \cite{Dioguardi:2013dd}.

The spin-lattice relaxation rate divided by temperature $(T_1T)^{-1}$ is shown in Fig.~\ref{fig:T1Tinv_layout} as a function of temperature for both nuclei. At high temperatures $(T_1T)^{-1}$ is roughly constant, indicating metallic Korringa behavior.  In the underdoped crystal $(T_1T)^{-1}$  goes through a peak at $T_N$  reflecting critical slowing down of the spin fluctuations.  In the optimally doped crystal $(T_1T)^{-1}$ continues to increase down to $T_c$.  These results are consistent with previously published data in polycrystalline samples \cite{Nakai:2013by,Nakai:2010ku}. The relaxation rates of the two nuclei scale roughly with one another, but there are important differences that emerge at low temperature, as discussed below.

The stretching exponent, $\beta$, shown in Fig. \ref{fig:T1Tinv_layout}, is a measure of the degree of dynamical inhomogeneity in the material \cite{Dioguardi:2013dd,Dioguardi:2015tx}.  $\beta = 1$ indicates homogeneous relaxation whereas $\beta<1$ indicates a distribution of local relaxation rates \cite{Johnston:2006gs}.  Both crystals and both sites become dynamically inhomogeneous below $\sim 100$ K, reaching down to $\beta = 0.6$ for the underdoped sample.  Similar behavior was observed in other iron pnictides \cite{hammerath:2010ki,hammerath:2013jj}.  Surprisingly, the degree of inhomogeneity does not appear to be reduced in the P-doped system as compared to the Co-doped system, despite the fact that the former is cleaner than the latter (comparisons of $\beta$ in \bafeasp\ and  BaFe$_{2-x}$Co$_x$As$_2$ are available in the supplemental material).

\begin{figure}
	\includegraphics[width=\linewidth]{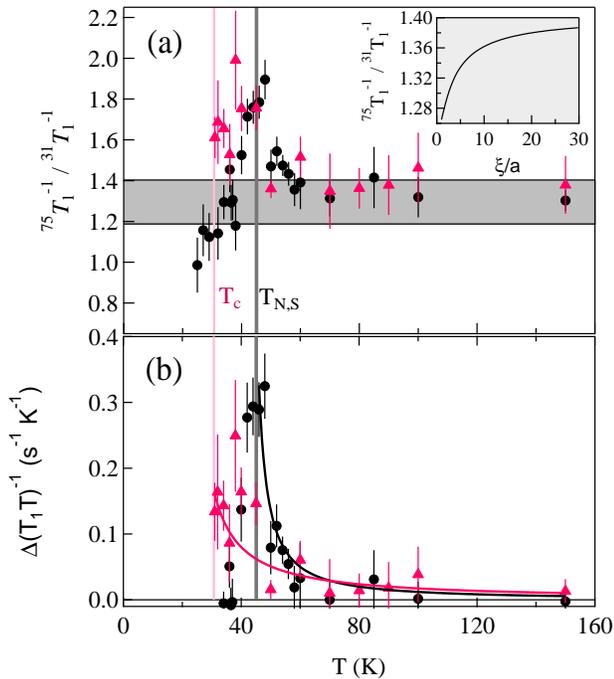}
	\caption{(a)Ratio of the spin lattice relaxation rates of As to P ($^{75}T_1^{-1}/ ^{31}T_1^{-1}$) and (b) $\Delta(T_1T)^{-1}$ vs. temperature for $x= 0.26$ ($\bullet$) and $x=0.33$ ($\blacktriangle$). The gray horizontal region indicates the range of values for purely magnetic fluctuations. INSET: The theoretical ratio  as a function of the antiferromagnetic correlation length, $\xi$.  Solid lines in (b) are best fits as discussed in the text. }
	\label{fig:T1inv_75As_div_31P}
\end{figure}

We now turn to the relationship between the As and the P relaxation rates. Fig.~\ref{fig:T1inv_75As_div_31P}(a) shows  $^{75}T_1^{-1}/^{31}T_1^{-1}$ as a function of temperature. This ratio is nearly constant and $\sim 1.3$ above approximately 60 K,  indicating a common relaxation mechanism for both sites.  However, below this temperature the ratio increases with decreasing temperature and reaches a maximum value of $\sim 2$ in both the underdoped and optimally doped crystals.  The strong temperature dependence of  this ratio reflects either a change in the antiferromagnetic fluctuations, or an additional relaxation mechanism present at the As site.

Spin fluctuations give rise to dynamical hyperfine fields causing nuclear spin relaxation. In order to properly extract the contribution of antiferromagnetic fluctuations to the relaxation rate, it is crucial to know the components of the hyperfine tensor, $\mathbb{B}$,  at both the As and the P. The hyperfine interaction is given by $\mathcal{H}_{hyp} = \sum_{i\in nn}\hat{\mathbf{I}}\cdot \mathbb{B}\cdot \mathbf{S}(\mathbf{r}_i)$,  where $\mathbf{S}(\mathbf{r}_i)$ is the electronic spin of the Fe, and $\hat{\mathbf{I}}$ is the nuclear spin of either the $^{31}$P or $^{75}$As \cite{takigawa2008}.   By comparing the Knight shift and magnetic susceptibility, Kitagawa \textit{et al.} found $^{75}B_{aa} = ^{75}\!\!B_{bb} = 0.66$ T/$\mu_B$, and $^{75}B_{cc} = 0.47$ T/$\mu_B$ \cite{takigawa2008}.   In \bafeasp, our measurements of the Knight shift (see supplemental material) indicate that $^{31}B_{aa}/^{75}B_{aa} = 0.40\pm 0.02$ in agreement with a previous study \cite{Nakai:2013by}.

\begin{figure}
	\includegraphics[width =\linewidth]{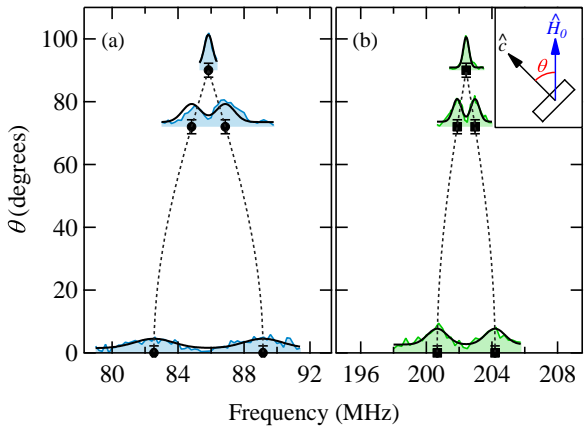}
	\caption{Spectra of the central resonance of the $^{75}$As (a) and $^{31}$P (b) as a function of angle $\theta$ with respect to the $\hat{c}$-axis for $x=0.26$. Markers represent the centers of the the respective peaks extracted from global fits (solid black lines) to each of three data sets at $0^\circ$, $72^\circ$, and $90^\circ$. Dashed black lines show the calculated angular dependence of the resonance centers of the peaks based on extracted fit coefficients for the internal fields.}
	\label{fig:Angular_Dep_Layout}
\end{figure}

It is also important to consider the off-diagonal component $B_{ac}$, which gives rise to the internal  field at the antiferromagnetic wavevector.  Previous measurements revealed that $^{75}B_{ac}= 0.43$ T/$\mu_B$ \cite{takigawa2008}.  To determine this component in \bafeasp\ we measured the angular dependence of the magnetic splitting of the central line in the antiferromagnetic phase of the underdoped sample.   Fig.~\ref{fig:Angular_Dep_Layout}  shows spectra of both the As and the P for various orientations of the crystal. The internal field, $\mathbf{H}_{int}$, is oriented along $\pm \hat{c}$-axis, giving rise to two separate resonances.
The P resonances are given by: 	$^{31}f = ~^{31}{\gamma}\left(1 + K\right)\left(H_0 \pm H_{int} \cos{\theta}\right)$, where $K$ is the Knight shift and $\theta$ is the angle between $\hat{c}$ and $\mathbf{H}_0$.
For the As, there is an additional shift due to the  quadrupolar interaction: $^{75}f=~^{75}\gamma\left(1 + K\right)\left(H_0 \pm H_{int} \cos{\theta}\right)+\Delta_Q (1 - 9 \cos^2{\theta})(1 - \cos^2{\theta})$, where $\Delta_Q = \frac{3}{64} e^2 Q^2 V_{zz}^{2}/ ^{75}\gamma H_0$, $Q$ is the quadrupolar moment and $V_{zz}$ is the component of the EFG tensor at the As \cite{slichter1990principles}.
Fitting  the angular dependent spectra, we extract internal fields $^{75}H_{int} = 0.45\pm 0.01$ T and $^{31}H_{int} = 0.100 \pm 0.001$ T, yielding $^{31}B_{ac}/^{75}B_{ac} = 0.226\pm 0.007$.  It is noteworthy that the transferred hyperfine couplings to the P are less than those to the As, which probably reflect the fact that the 4p orbitals at the As are more extended.  Previous studies of the hyperfine couplings at the As and P sites in other compounds have found a similar ratio \cite{Kinouchi2012}.

With the knowledge of the hyperfine couplings, it is now possible to compute the magnetic component of the spin-lattice relaxation rate:
\begin{equation}
	\label{eqn:dynamical_susceptibility}
	W_0 = \gamma^2 k_B T \lim_{\omega \rightarrow 0} \sum\limits_{\mathbf{q},\alpha,\beta} \mathcal{F}_{\alpha\beta}(\mathbf{q}) \frac{\textrm{Im}\chi_{m}^{\alpha\beta}(\mathbf{q},\omega)}{\hslash \omega},
\end{equation}
where $\mathcal{F}_{\alpha\beta}(\mathbf{q})$ are the form factors (given in the Supplemental Material), $\chi^{\alpha\beta}_{m}(\mathbf{q},\omega)$ is  the dynamical magnetic susceptibility, and $\alpha,\beta = \left\{ a,b,c \right\}$ \cite{T1formfactorsArsenides}.  For purely magnetic fluctuations, $T_{1}^{-1} = 2W_0$.
Because the hyperfine coupling ratios  $^{31}B_{aa}/^{75}B_{aa}$ and  $^{31}B_{ac}/^{75}B_{ac}$ are not the same, the form factors $\mathcal{F}_{\alpha\beta}(\mathbf{q})$ for the two sites do not simply scale with one another.  As a result, the ratio $^{75}T_1^{-1}/^{31}T_1^{-1}$ will depend of the detailed $\mathbf{q}$-dependence of $\chi_m^{\alpha\beta}(\mathbf{q},\omega)$, which can change with temperature. To estimate the effect of antiferromagnetic correlations on the \slrr\ ratio, we use Eq. \ref{eqn:dynamical_susceptibility} and the dynamical magnetic susceptibility:
\begin{equation}
\chi^{\alpha\alpha}_m(\mathbf{q},\omega) = \sum_{j=1,2}\frac{\chi_{\alpha\alpha}(\mathbf{Q}_j)}{1+|\mathbf{q} - \mathbf{Q}_j|^2\xi^2 -i\omega/\omega_{sf}},
\label{eqn:chiMMP}
\end{equation}
where $\xi$ is the antiferromagnetic correlation length, $\omega_{sf}$ is the characteristic spin fluctuation frequency, and $\chi_{\alpha\alpha}(\mathbf{Q}_j)$ is the value of the susceptibility at the ordering wavevectors $\mathbf{Q}_1 = \{\pi/a,0\}$ and $\mathbf{Q}_2 = \{0,\pi/a\}$ \cite{StrainedPnictidesNS2014science,MMPT1inYBCO}.
The inset of Fig. \ref{fig:T1inv_75As_div_31P}(a) shows the calculated \slrr\ ratio as a function of correlation length, with the following assumptions: (i) $^{31}B_{cc}/^{75}B_{cc} = ^{31}\!\!B_{aa}/^{75}B_{aa}$, (ii) $B_{ab}$ is negligible, and (iii) $\chi_{\alpha\alpha}(\mathbf{Q}_i)$ is the same for all values of $i$ and $\alpha$ (isotropic fluctuations).  This quantity changes only slightly with $\xi$, varying between 1.19 ($\xi=0$) to 1.40 ($\xi=\infty$), as shown by the gray area in Fig. \ref{fig:T1inv_75As_div_31P}(a).  For the underdoped sample, the experimental ratio exceeds this prediction below the structural transition, reaching up to $\sim 2$ at $T_N$.  The experimental ratio for the optimally doped sample reaches the same value just above $T_c$.  It is clear that magnetic fluctuations alone cannot explain this large increase, suggesting that there is an additional relaxation channel affecting the As site.

Since the $^{75}$As has spin $I=3/2$ it is susceptible to relaxation by fluctuations of the EFG through the quadrupolar coupling. Quadrupolar spin-lattice relaxation of nuclear spins with $I>\frac{1}{2}$ is described by the Hamiltonian:
\begin{equation}
\mathcal{H}_Q(t) =  \frac{eQ}{4I (2I-1)}\sum_{k=-2}^{2}V_k(t)\hat{T}_{2k},
\end{equation}
where $\hat{T}_{2k}$ are the spherical tensor operators,  $V_{0} = V_{zz}$, $V_{\pm 1} = V_{zx}\pm i V_{zy}$,  $V_{\pm 2} = \frac{1}{2}\left(V_{xx} - V_{yy}\right) \pm i V_{xy}$, and $V_{\alpha\beta}$ are the components of the EFG tensor \cite{slichter1990principles}.  The simultaneous presence of both magnetic and quadrupolar fluctuations has been discussed in detail by Suter, who  has shown that these fluctuations give rise to three relaxation channels for the nuclear spins, one purely magnetic and two quadrupolar \cite{Suter:1999kw}. The magnetic relaxation channel is described by Eq. \ref{eqn:dynamical_susceptibility}, and the quadrupolar relaxation rates are given by: $W_{1,2} = \left(\frac{eQ}{\hbar}\right)^2 \int_{-\infty}^{\infty} \langle V_{+1,2}(\tau)V_{-1,2}(0)\rangle e^{i\omega\tau}d\tau$. There are two components to the EFG at the As, one from the lattice orthorhombic distortion and the other from unequal populations of the As $4p_{x,y}$ orbitals, such that $V_{\pm1} = 0$ and $V_{\pm2} = V_{\pm2}^{lat} + V_{\pm2}^{4p}$ \cite{IshidaPdopedBa122JPSJ2012}.  From the definition, we see that $\left(V_{+2}+V_{-2}\right)$ has $B_{2g}$ symmetry,
whereas $i\left(V_{+2}-V_{-2}\right)$ has $B_{1g}$ symmetry (in
the coordinate system of the crystallographic tetragonal unit cell). Thus, the
former couples directly to the nematic order parameter. Using the fluctuation-dissipation theorem, we can express the quadrupolar relaxation rate in terms the of the dynamical nematic susceptibility $\chi_n$:
\begin{equation}
	\label{eqn:Moriya_nematic}
	W_2 =  \left(\frac{eQ}{\hbar}\right)^2 k_B T \lim_{\omega \rightarrow 0} \sum\limits_{\mathbf{q}} \frac{\mathrm{Im}\chi_{n}(\mathbf{q},\omega)}{\hslash \omega}.
\end{equation}
Note that $\chi_n$ is defined in terms of the EFG, and that there is no form factor for the nematic fluctuations because  the on-site orbital occupations are the dominant contribution to the EFG.   The nematic fluctuations order at $\mathbf{q}=(0,0)$ with Curie-Weiss behavior, and the existence of a Fermi surface implies Landau damping \cite{Chu:2012ez,IsingSpinOrderSachdev2008}.
We estimate the magnitude of $W_2$ by considering the static EFG at the As site. In the orthorhombic phase, $V_{\pm2}$ develops a finite value reflecting the static nematic order \cite{takigawa2008}. If we assume that the lattice, orbital and spin degrees of freedom have a similar power spectrum,  then ratio of the quadrupolar to magnetic relaxation rates is  $W_2/W_0 \sim \left(eQ\overline{V}_{2}/{\gamma \hbar \overline{h}}\right)^2$, where $\overline{V}_{2}$ and $\overline{h}$ are the root mean square of the EFG and hyperfine field fluctuations, respectively. Previous field-dependent studies in {BaFe$_{2-x}$Co$_x$As$_2$} found $\overline{h}\sim 40$ G \cite{Dioguardi:2015tx}.  $eQ\overline{V}_{2}/\hbar$ reaches a static value of 2.6 MHz in the orthorhombic phase of the parent compound \cite{takigawa2008},  and a value of $\sim 0.3$ MHz at the structural transition in BaFe$_2$(As$_{0.96}$P$_{0.04}$)$_2$ \cite{IshidaPdopedBa122JPSJ2012}. We estimate $W_2/W_0$ can reach a maximum of $\sim 8.4$ at $T_s = T_N$, thus it is clear that both magnetic and nematic fluctuations are of comparable magnitude and can contribute to the spin-lattice relaxation of the As.  We conclude that the enhanced  temperature dependent ratio seen in Fig. \ref{fig:T1inv_75As_div_31P}(a) reflects the presence of nematic fluctuations.

Note that in the presence of both quadrupolar and magnetic fluctuations, the two relaxation rates $W_0$ and $W_2$ become entangled and the exact form  of the magnetization recovery becomes complex \cite{Suter:1999kw}.   Attempts to fit the recovery data to a modified relaxation form with $W_2$ as a floating parameter do not necessarily lead to a better quality of fit as measured by the $\chi^2$ value.  Such fits have poor precision because both relaxation rates $W_0$ and $W_2$ follow a broad distribution function, thus the relaxation curve is stretched.  This distribution is evident in the P relaxation, which has no quadrupolar relaxation channel but still exhibits stretched recovery.  Thus the difference $\Delta(T_1T)^{-1} = ^{75}(T_1T)^{-1} - \kappa(^{31}T_1T)^{-1} $, where $\kappa = 1.31$ is the high temperature ratio of the As to P relaxation rates is not simply proportional to $W_{2}$. Yet, in order to assess qualitatively the contribution from $W_{2}$,
we can still focus on this quantity, plotted in Fig. \ref{fig:T1inv_75As_div_31P}(b),
since $W_{2}=0$ would imply $\Delta(T_{1}T)^{-1}=0$.

It is clear from Fig. \ref{fig:T1inv_75As_div_31P}(b) that nematic fluctuations are present in both the underdoped and optimally-doped samples. In the underdoped crystal, the tetragonal-orthorhombic phase transition $T_s$ coincides with $T_N$, and the nematic fluctuations diverge at this phase transition \cite{PBa122StructureMagneticPRL2015}. The solid lines through the data points are best fits to the expression $\Delta(T_1T)^{-1} = A/(T-T_0)^{n}$, where for the underdoped crystal $n=1.4\pm1.1$ and $T_0 = 41.5\pm 5.7$ K.
In the optimally doped sample, the data reveal that nematic fluctuations are present in the tetragonal phase down to $T_c$. The best fit through the data points yields $n=1.5\pm 1.9$ and $T_0 = 13\pm 33$ K.
Although the exact relationship between $\Delta(T_{1}T)^{-1}$ and $W_{2}\propto\chi_{n}$
is not known, it is interesting to note that this result is consistent with elastoresistance measurements, which indicate an enhanced $\chi_n$ near a putative quantum critical point \cite{Kuo2015}.

An alternative explanation for the behavior of the \slrr\ ratio in Fig. \ref{fig:T1inv_75As_div_31P}(a) is that the spin fluctuations are locally suppressed at the P sites.  In this case, the As relaxation would not be enhanced by nematic fluctuations, but rather the P relaxation rate would be suppressed.  A recent comparison of NMR of $^{63}$Cu and $^{75}$As in Ba(Fe$_{1-x}$Cu$_x$)$_2$As$_2$ revealed a slightly reduced relaxation at the Cu site \cite{Takeda:2014ia}.  However, the form factor and hyperfine couplings to the Cu are different than the As, which could explain the difference. Furthermore, since $^{63}Q < ^{75}Q$, it is possible that this difference reflects the reduced quadrupolar interaction at the Cu sites. Thus, our
results indicate that $^{75}\left(T_{1}T\right)^{-1}$ always contains
a contribution from nematic fluctuations. The extent to which these
fluctuations affect the expected Curie-Weiss behavior $^{75}\left(T_{1}T\right)^{-1}\propto\left(T-T_{N}\right)^{-1}$
near a magnetic transition remains to be further investigated. For
instance, in systems with split nematic and magnetic transitions,
such as NaFeAs and Ba(Fe$_{1-x}$Co$_{x}$)$_{2}$As$_{2}$, $^{75}\left(T_{1}T\right)^{-1}$
seems to display no additional peaks at $T_{s}$, suggesting that
$W_{2}$ may be subleading compared to $W_{0}$, at least in those
materials.

The approach we have taken using the susceptibility in Eq. \ref{eqn:chiMMP} to estimate the magnetic contribution to the relaxation is essentially identical to a recent study using self-consistent renormalization spin-fluctuation theory (SCR) \cite{Nakai:2013by}.  In the previous study, the authors found that the spin fluctuations evolve with doping and exhibit quantum critical behavior near optimal doping for superconductivity. These antiferromagnetic spin fluctuations may provide the pairing glue for the superconductivity \cite{chubukovreview}. Our results indicate that  these spin fluctuations are accompanied by nematic fluctuations in the optimally doped sample.  It is possible that the nematic fluctuations may also be important for the superconducting mechanism \cite{KivelsonNematicQCP2015}.

It is interesting to consider why the inhomogeneous glassy behavior is unaffected by isovalent P doping on the As site rather than electron doping at the Fe site.  Elastoresistance measurements of the nematic susceptibility find Curie-Weiss behavior over a broad range of temperatures in various doped iron pnictides \cite{Kuo2015}.  However, at low temperatures in both the electron and hole doped systems, the nematic susceptibility exhibits a deviation from Curie-Weiss behavior that may arise from quenched disorder.  The P-doped system, on the other hand, showed no such deviation suggesting that this system contains the least amount of disorder.  Our NMR results show no  distinction and indicate a similar glassy distribution of relaxation rates in P, Co and Cu doped systems.  It is likely that doping at the As site still disrupts the exchange interaction between the Fe orbitals, providing  a source of frustration. Alternatively, these results could also indicate that
the electronic glassiness is driven by frustrated interactions rather
than chemical inhomogeneity \cite{Schmalian_Wolynes00}.

To summarize, we have measured the hyperfine couplings and spin lattice relaxation rates of the As and P sites in \bafeasp.  We find that spin fluctuations alone are insufficient to explain the ratio between the relaxation rates at these two sites, however critical slowing down of nematic fluctuations in the tetragonal phase that couple to the quadrupolar moment of the As can explain the enhanced relaxation at the As site. In contrast to torque magnetometry and optical measurements, our results show no evidence for a phase transition above $T_s$, however the critical fluctuations persist  well above the structural transition \cite{Kasahara2012,Thewalt2015}. Furthermore, the presence of inhomogeneous strain distributions in the tetragonal phase may be responsible for the distribution of relaxation rates that we observe.

We thank P. Canfield, S. Bud'ko, S. Kivelson, J. Schmalian, P. Carreta, and E. Carlson for stimulating discussions. Work at UC Davis  was supported by the UC Lab Research Fee Program and the NNSA under the Stewardship Science Academic Alliances program through U.S. DOE Research Grant No.\ DE-FG52-09NA29464. Work at Stanford University was supported by the DOE, Office of Basic Energy Sciences, under Contract No. DE-AC02-76SF00515. RMF is supported by the U.S. Department of Energy, Office of Science, Basic Energy Sciences, under award number DE-SC0012336.


\input{P-doped_Ba122_PRL.bbl}

\end{document}

%% file: P-doped_Ba122_PRL.bbl
%

%% file: P-doped_Ba122_PRL.bbl
\begin{thebibliography}{45}%
\makeatletter
\providecommand \@ifxundefined [1]{%
 \@ifx{#1\undefined}
}%
\providecommand \@ifnum [1]{%
 \ifnum #1\expandafter \@firstoftwo
 \else \expandafter \@secondoftwo
 \fi
}%
\providecommand \@ifx [1]{%
 \ifx #1\expandafter \@firstoftwo
 \else \expandafter \@secondoftwo
 \fi
}%
\providecommand \natexlab [1]{#1}%
\providecommand \enquote  [1]{``#1''}%
\providecommand \bibnamefont  [1]{#1}%
\providecommand \bibfnamefont [1]{#1}%
\providecommand \citenamefont [1]{#1}%
\providecommand \href@noop [0]{\@secondoftwo}%
\providecommand \href [0]{\begingroup \@sanitize@url \@href}%
\providecommand \@href[1]{\@@startlink{#1}\@@href}%
\providecommand \@@href[1]{\endgroup#1\@@endlink}%
\providecommand \@sanitize@url [0]{\catcode `\\12\catcode `\$12\catcode
  `\&12\catcode `\#12\catcode `\^12\catcode `\_12\catcode `\%12\relax}%
\providecommand \@@startlink[1]{}%
\providecommand \@@endlink[0]{}%
\providecommand \url  [0]{\begingroup\@sanitize@url \@url }%
\providecommand \@url [1]{\endgroup\@href {#1}{\urlprefix }}%
\providecommand \urlprefix  [0]{URL }%
\providecommand \Eprint [0]{\href }%
\providecommand \doibase [0]{http://dx.doi.org/}%
\providecommand \selectlanguage [0]{\@gobble}%
\providecommand \bibinfo  [0]{\@secondoftwo}%
\providecommand \bibfield  [0]{\@secondoftwo}%
\providecommand \translation [1]{[#1]}%
\providecommand \BibitemOpen [0]{}%
\providecommand \bibitemStop [0]{}%
\providecommand \bibitemNoStop [0]{.\EOS\space}%
\providecommand \EOS [0]{\spacefactor3000\relax}%
\providecommand \BibitemShut  [1]{\csname bibitem#1\endcsname}%
\let\auto@bib@innerbib\@empty
\bibitem [{\citenamefont {Johnston}(2010)}]{Johnston:2010cs}%
  \BibitemOpen
  \bibfield  {author} {\bibinfo {author} {\bibfnamefont {D.~C.}\ \bibnamefont
  {Johnston}},\ }\href {\doibase 10.1080/00018732.2010.513480} {\bibfield
  {journal} {\bibinfo  {journal} {Adv. Phys.}\ }\textbf {\bibinfo {volume}
  {59}},\ \bibinfo {pages} {803} (\bibinfo {year} {2010})}\BibitemShut
  {NoStop}%
\bibitem [{\citenamefont {Moriya}\ and\ \citenamefont
  {Ueda}(2003)}]{uedareview}%
  \BibitemOpen
  \bibfield  {author} {\bibinfo {author} {\bibfnamefont {T.}~\bibnamefont
  {Moriya}}\ and\ \bibinfo {author} {\bibfnamefont {K.}~\bibnamefont {Ueda}},\
  }\href@noop {} {\bibfield  {journal} {\bibinfo  {journal} {Rep. Prog. Phys.}\
  }\textbf {\bibinfo {volume} {66}},\ \bibinfo {pages} {1299} (\bibinfo {year}
  {2003})}\BibitemShut {NoStop}%
\bibitem [{\citenamefont {Chubukov}(2012)}]{chubukovreview}%
  \BibitemOpen
  \bibfield  {author} {\bibinfo {author} {\bibfnamefont {A.}~\bibnamefont
  {Chubukov}},\ }\href {\doibase 10.1146/annurev-conmatphys-020911-125055}
  {\bibfield  {journal} {\bibinfo  {journal} {Annual Review of Condensed Matter
  Physics}\ }\textbf {\bibinfo {volume} {3}},\ \bibinfo {pages} {57} (\bibinfo
  {year} {2012})}\BibitemShut {NoStop}%
\bibitem [{\citenamefont {Fang}\ \emph {et~al.}(2008)\citenamefont {Fang},
  \citenamefont {Yao}, \citenamefont {Tsai}, \citenamefont {Hu},\ and\
  \citenamefont {Kivelson}}]{Fang:2008by}%
  \BibitemOpen
  \bibfield  {author} {\bibinfo {author} {\bibfnamefont {C.}~\bibnamefont
  {Fang}}, \bibinfo {author} {\bibfnamefont {H.}~\bibnamefont {Yao}}, \bibinfo
  {author} {\bibfnamefont {W.-F.}\ \bibnamefont {Tsai}}, \bibinfo {author}
  {\bibfnamefont {J.}~\bibnamefont {Hu}}, \ and\ \bibinfo {author}
  {\bibfnamefont {S.~A.}\ \bibnamefont {Kivelson}},\ }\href {\doibase
  10.1103/PhysRevB.77.224509} {\bibfield  {journal} {\bibinfo  {journal} {Phys.
  Rev. B}\ }\textbf {\bibinfo {volume} {77}},\ \bibinfo {pages} {224509}
  (\bibinfo {year} {2008})}\BibitemShut {NoStop}%
\bibitem [{\citenamefont {Xu}\ \emph {et~al.}(2008)\citenamefont {Xu},
  \citenamefont {M\"uller},\ and\ \citenamefont
  {Sachdev}}]{IsingSpinOrderSachdev2008}%
  \BibitemOpen
  \bibfield  {author} {\bibinfo {author} {\bibfnamefont {C.}~\bibnamefont
  {Xu}}, \bibinfo {author} {\bibfnamefont {M.}~\bibnamefont {M\"uller}}, \ and\
  \bibinfo {author} {\bibfnamefont {S.}~\bibnamefont {Sachdev}},\ }\href
  {\doibase 10.1103/PhysRevB.78.020501} {\bibfield  {journal} {\bibinfo
  {journal} {Phys. Rev. B}\ }\textbf {\bibinfo {volume} {78}},\ \bibinfo
  {pages} {020501} (\bibinfo {year} {2008})}\BibitemShut {NoStop}%
\bibitem [{\citenamefont {Fernandes}\ \emph {et~al.}(2014)\citenamefont
  {Fernandes}, \citenamefont {Chubukov},\ and\ \citenamefont
  {Schmalian}}]{Fernandes:2014en}%
  \BibitemOpen
  \bibfield  {author} {\bibinfo {author} {\bibfnamefont {R.~M.}\ \bibnamefont
  {Fernandes}}, \bibinfo {author} {\bibfnamefont {A.~V.}\ \bibnamefont
  {Chubukov}}, \ and\ \bibinfo {author} {\bibfnamefont {J.}~\bibnamefont
  {Schmalian}},\ }\href {\doibase 10.1038/nphys2877} {\bibfield  {journal}
  {\bibinfo  {journal} {Nature Physics}\ }\textbf {\bibinfo {volume} {10}},\
  \bibinfo {pages} {97} (\bibinfo {year} {2014})}\BibitemShut {NoStop}%
\bibitem [{\citenamefont {Gallais}\ \emph {et~al.}(2013)\citenamefont
  {Gallais}, \citenamefont {Fernandes}, \citenamefont {Paul}, \citenamefont
  {Chauvi{\`e}re}, \citenamefont {Yang}, \citenamefont {M{\'e}asson},
  \citenamefont {Cazayous}, \citenamefont {Sacuto}, \citenamefont {Colson},\
  and\ \citenamefont {Forget}}]{Gallais:2013fj}%
  \BibitemOpen
  \bibfield  {author} {\bibinfo {author} {\bibfnamefont {Y.}~\bibnamefont
  {Gallais}}, \bibinfo {author} {\bibfnamefont {R.~M.}\ \bibnamefont
  {Fernandes}}, \bibinfo {author} {\bibfnamefont {I.}~\bibnamefont {Paul}},
  \bibinfo {author} {\bibfnamefont {L.}~\bibnamefont {Chauvi{\`e}re}}, \bibinfo
  {author} {\bibfnamefont {Y.~X.}\ \bibnamefont {Yang}}, \bibinfo {author}
  {\bibfnamefont {M.~A.}\ \bibnamefont {M{\'e}asson}}, \bibinfo {author}
  {\bibfnamefont {M.}~\bibnamefont {Cazayous}}, \bibinfo {author}
  {\bibfnamefont {A.}~\bibnamefont {Sacuto}}, \bibinfo {author} {\bibfnamefont
  {D.}~\bibnamefont {Colson}}, \ and\ \bibinfo {author} {\bibfnamefont
  {A.}~\bibnamefont {Forget}},\ }\href {\doibase
  10.1103/PhysRevLett.111.267001} {\bibfield  {journal} {\bibinfo  {journal}
  {Phys. Rev. Lett.}\ }\textbf {\bibinfo {volume} {111}},\ \bibinfo {pages}
  {267001} (\bibinfo {year} {2013})}\BibitemShut {NoStop}%
\bibitem [{\citenamefont {Chu}\ \emph {et~al.}(2010)\citenamefont {Chu},
  \citenamefont {Analytis}, \citenamefont {Press}, \citenamefont {De~Greve},
  \citenamefont {Ladd}, \citenamefont {Yamamoto},\ and\ \citenamefont
  {Fisher}}]{Chu:2010hg}%
  \BibitemOpen
  \bibfield  {author} {\bibinfo {author} {\bibfnamefont {J.-H.}\ \bibnamefont
  {Chu}}, \bibinfo {author} {\bibfnamefont {J.~G.}\ \bibnamefont {Analytis}},
  \bibinfo {author} {\bibfnamefont {D.}~\bibnamefont {Press}}, \bibinfo
  {author} {\bibfnamefont {K.}~\bibnamefont {De~Greve}}, \bibinfo {author}
  {\bibfnamefont {T.~D.}\ \bibnamefont {Ladd}}, \bibinfo {author}
  {\bibfnamefont {Y.}~\bibnamefont {Yamamoto}}, \ and\ \bibinfo {author}
  {\bibfnamefont {I.~R.}\ \bibnamefont {Fisher}},\ }\href {\doibase
  10.1103/PhysRevB.81.214502} {\bibfield  {journal} {\bibinfo  {journal} {Phys.
  Rev. B}\ }\textbf {\bibinfo {volume} {81}},\ \bibinfo {pages} {214502}
  (\bibinfo {year} {2010})}\BibitemShut {NoStop}%
\bibitem [{\citenamefont {Chu}\ \emph {et~al.}(2012)\citenamefont {Chu},
  \citenamefont {Kuo}, \citenamefont {Analytis},\ and\ \citenamefont
  {Fisher}}]{Chu:2012ez}%
  \BibitemOpen
  \bibfield  {author} {\bibinfo {author} {\bibfnamefont {J.~H.}\ \bibnamefont
  {Chu}}, \bibinfo {author} {\bibfnamefont {H.~H.}\ \bibnamefont {Kuo}},
  \bibinfo {author} {\bibfnamefont {J.~G.}\ \bibnamefont {Analytis}}, \ and\
  \bibinfo {author} {\bibfnamefont {I.~R.}\ \bibnamefont {Fisher}},\ }\href
  {\doibase 10.1126/science.1221713} {\bibfield  {journal} {\bibinfo  {journal}
  {Science}\ }\textbf {\bibinfo {volume} {337}},\ \bibinfo {pages} {710}
  (\bibinfo {year} {2012})}\BibitemShut {NoStop}%
\bibitem [{\citenamefont {Kuo}\ \emph {et~al.}(2015)\citenamefont {Kuo},
  \citenamefont {Chu}, \citenamefont {Kivelson},\ and\ \citenamefont
  {Fisher}}]{Kuo2015}%
  \BibitemOpen
  \bibfield  {author} {\bibinfo {author} {\bibfnamefont {H.-H.}\ \bibnamefont
  {Kuo}}, \bibinfo {author} {\bibfnamefont {J.-H.}\ \bibnamefont {Chu}},
  \bibinfo {author} {\bibfnamefont {S.~A.}\ \bibnamefont {Kivelson}}, \ and\
  \bibinfo {author} {\bibfnamefont {I.~R.}\ \bibnamefont {Fisher}},\
  }\href@noop {} {\  (\bibinfo {year} {2015})},\ \Eprint
  {http://arxiv.org/abs/1503.00402} {1503.00402} \BibitemShut {NoStop}%
\bibitem [{\citenamefont {B{\"o}hmer}\ \emph {et~al.}(2014)\citenamefont
  {B{\"o}hmer}, \citenamefont {Burger}, \citenamefont {Hardy}, \citenamefont
  {Wolf}, \citenamefont {Schweiss}, \citenamefont {Fromknecht}, \citenamefont
  {Reinecker}, \citenamefont {Schranz},\ and\ \citenamefont
  {Meingast}}]{Bohmer:2014jv}%
  \BibitemOpen
  \bibfield  {author} {\bibinfo {author} {\bibfnamefont {A.~E.}\ \bibnamefont
  {B{\"o}hmer}}, \bibinfo {author} {\bibfnamefont {P.}~\bibnamefont {Burger}},
  \bibinfo {author} {\bibfnamefont {F.}~\bibnamefont {Hardy}}, \bibinfo
  {author} {\bibfnamefont {T.}~\bibnamefont {Wolf}}, \bibinfo {author}
  {\bibfnamefont {P.}~\bibnamefont {Schweiss}}, \bibinfo {author}
  {\bibfnamefont {R.}~\bibnamefont {Fromknecht}}, \bibinfo {author}
  {\bibfnamefont {M.}~\bibnamefont {Reinecker}}, \bibinfo {author}
  {\bibfnamefont {W.}~\bibnamefont {Schranz}}, \ and\ \bibinfo {author}
  {\bibfnamefont {C.}~\bibnamefont {Meingast}},\ }\href {\doibase
  10.1103/PhysRevLett.112.047001} {\bibfield  {journal} {\bibinfo  {journal}
  {Phys. Rev. Lett.}\ }\textbf {\bibinfo {volume} {112}},\ \bibinfo {pages}
  {047001} (\bibinfo {year} {2014})}\BibitemShut {NoStop}%
\bibitem [{\citenamefont {Kretzschmar}\ \emph {et~al.}(2015)\citenamefont
  {Kretzschmar}, \citenamefont {B\"ohm}, \citenamefont {Karahasanovi\'c},
  \citenamefont {Muschler}, \citenamefont {Baum}, \citenamefont {Jost},
  \citenamefont {Schmalian}, \citenamefont {Caprara}, \citenamefont {Grilli},
  \citenamefont {Castro}, \citenamefont {Analytis}, \citenamefont {Chu},
  \citenamefont {Fisher},\ and\ \citenamefont {Hackl}}]{Kretzschmar2015}%
  \BibitemOpen
  \bibfield  {author} {\bibinfo {author} {\bibfnamefont {F.}~\bibnamefont
  {Kretzschmar}}, \bibinfo {author} {\bibfnamefont {T.}~\bibnamefont {B\"ohm}},
  \bibinfo {author} {\bibfnamefont {U.}~\bibnamefont {Karahasanovi\'c}},
  \bibinfo {author} {\bibfnamefont {B.}~\bibnamefont {Muschler}}, \bibinfo
  {author} {\bibfnamefont {A.}~\bibnamefont {Baum}}, \bibinfo {author}
  {\bibfnamefont {D.}~\bibnamefont {Jost}}, \bibinfo {author} {\bibfnamefont
  {J.}~\bibnamefont {Schmalian}}, \bibinfo {author} {\bibfnamefont
  {S.}~\bibnamefont {Caprara}}, \bibinfo {author} {\bibfnamefont
  {M.}~\bibnamefont {Grilli}}, \bibinfo {author} {\bibfnamefont {C.~D.}\
  \bibnamefont {Castro}}, \bibinfo {author} {\bibfnamefont {J.~G.}\
  \bibnamefont {Analytis}}, \bibinfo {author} {\bibfnamefont {J.-H.}\
  \bibnamefont {Chu}}, \bibinfo {author} {\bibfnamefont {I.~R.}\ \bibnamefont
  {Fisher}}, \ and\ \bibinfo {author} {\bibfnamefont {R.}~\bibnamefont
  {Hackl}},\ }\href@noop {} {\  (\bibinfo {year} {2015})},\ \Eprint
  {http://arxiv.org/abs/1507.06116} {1507.06116} \BibitemShut {NoStop}%
\bibitem [{\citenamefont {Gallais}\ and\ \citenamefont
  {Paul}(2015)}]{Gallais2015}%
  \BibitemOpen
  \bibfield  {author} {\bibinfo {author} {\bibfnamefont {Y.}~\bibnamefont
  {Gallais}}\ and\ \bibinfo {author} {\bibfnamefont {I.}~\bibnamefont {Paul}},\
  }\href@noop {} {\  (\bibinfo {year} {2015})},\ \Eprint
  {http://arxiv.org/abs/1508.01319} {1508.01319} \BibitemShut {NoStop}%
\bibitem [{\citenamefont {Thorsmoelle}\ \emph {et~al.}(2014)\citenamefont
  {Thorsmoelle}, \citenamefont {Khodas}, \citenamefont {Yin}, \citenamefont
  {Zhang}, \citenamefont {Carr}, \citenamefont {Dai},\ and\ \citenamefont
  {Blumberg}}]{Thorsmoelle2014}%
  \BibitemOpen
  \bibfield  {author} {\bibinfo {author} {\bibfnamefont {V.~K.}\ \bibnamefont
  {Thorsmoelle}}, \bibinfo {author} {\bibfnamefont {M.}~\bibnamefont {Khodas}},
  \bibinfo {author} {\bibfnamefont {Z.~P.}\ \bibnamefont {Yin}}, \bibinfo
  {author} {\bibfnamefont {C.}~\bibnamefont {Zhang}}, \bibinfo {author}
  {\bibfnamefont {S.~V.}\ \bibnamefont {Carr}}, \bibinfo {author}
  {\bibfnamefont {P.}~\bibnamefont {Dai}}, \ and\ \bibinfo {author}
  {\bibfnamefont {G.}~\bibnamefont {Blumberg}},\ }\href@noop {} {\  (\bibinfo
  {year} {2014})},\ \Eprint {http://arxiv.org/abs/1410.6456} {1410.6456}
  \BibitemShut {NoStop}%
\bibitem [{\citenamefont {Lederer}\ \emph {et~al.}(2015)\citenamefont
  {Lederer}, \citenamefont {Schattner}, \citenamefont {Berg},\ and\
  \citenamefont {Kivelson}}]{KivelsonNematicQCP2015}%
  \BibitemOpen
  \bibfield  {author} {\bibinfo {author} {\bibfnamefont {S.}~\bibnamefont
  {Lederer}}, \bibinfo {author} {\bibfnamefont {Y.}~\bibnamefont {Schattner}},
  \bibinfo {author} {\bibfnamefont {E.}~\bibnamefont {Berg}}, \ and\ \bibinfo
  {author} {\bibfnamefont {S.~A.}\ \bibnamefont {Kivelson}},\ }\href {\doibase
  10.1103/PhysRevLett.114.097001} {\bibfield  {journal} {\bibinfo  {journal}
  {Phys. Rev. Lett.}\ }\textbf {\bibinfo {volume} {114}},\ \bibinfo {pages}
  {097001} (\bibinfo {year} {2015})}\BibitemShut {NoStop}%
\bibitem [{\citenamefont {Maier}\ and\ \citenamefont
  {Scalapino}(2014)}]{ScalapinoNematicQCP2015}%
  \BibitemOpen
  \bibfield  {author} {\bibinfo {author} {\bibfnamefont {T.~A.}\ \bibnamefont
  {Maier}}\ and\ \bibinfo {author} {\bibfnamefont {D.~J.}\ \bibnamefont
  {Scalapino}},\ }\href {\doibase 10.1103/PhysRevB.90.174510} {\bibfield
  {journal} {\bibinfo  {journal} {Phys. Rev. B}\ }\textbf {\bibinfo {volume}
  {90}},\ \bibinfo {pages} {174510} (\bibinfo {year} {2014})}\BibitemShut
  {NoStop}%
\bibitem [{\citenamefont {Fernandes}\ \emph {et~al.}(2013)\citenamefont
  {Fernandes}, \citenamefont {B{\"o}hmer}, \citenamefont {Meingast},\ and\
  \citenamefont {Schmalian}}]{Fernandes:2013ib}%
  \BibitemOpen
  \bibfield  {author} {\bibinfo {author} {\bibfnamefont {R.~M.}\ \bibnamefont
  {Fernandes}}, \bibinfo {author} {\bibfnamefont {A.~E.}\ \bibnamefont
  {B{\"o}hmer}}, \bibinfo {author} {\bibfnamefont {C.}~\bibnamefont
  {Meingast}}, \ and\ \bibinfo {author} {\bibfnamefont {J.}~\bibnamefont
  {Schmalian}},\ }\href {\doibase 10.1103/PhysRevLett.111.137001} {\bibfield
  {journal} {\bibinfo  {journal} {Phys. Rev. Lett.}\ }\textbf {\bibinfo
  {volume} {111}},\ \bibinfo {pages} {137001} (\bibinfo {year}
  {2013})}\BibitemShut {NoStop}%
\bibitem [{\citenamefont {B{\"o}hmer}\ and\ \citenamefont
  {Meingast}(2015)}]{bohmer2015electronic}%
  \BibitemOpen
  \bibfield  {author} {\bibinfo {author} {\bibfnamefont {A.~E.}\ \bibnamefont
  {B{\"o}hmer}}\ and\ \bibinfo {author} {\bibfnamefont {C.}~\bibnamefont
  {Meingast}},\ }\href@noop {} {\bibfield  {journal} {\bibinfo  {journal}
  {arXiv preprint arXiv:1505.05120}\ } (\bibinfo {year} {2015})}\BibitemShut
  {NoStop}%
\bibitem [{\citenamefont {Lu}\ \emph {et~al.}(2014)\citenamefont {Lu},
  \citenamefont {Park}, \citenamefont {Zhang}, \citenamefont {Luo},
  \citenamefont {Nevidomskyy}, \citenamefont {Si},\ and\ \citenamefont
  {Dai}}]{StrainedPnictidesNS2014science}%
  \BibitemOpen
  \bibfield  {author} {\bibinfo {author} {\bibfnamefont {X.}~\bibnamefont
  {Lu}}, \bibinfo {author} {\bibfnamefont {J.~T.}\ \bibnamefont {Park}},
  \bibinfo {author} {\bibfnamefont {R.}~\bibnamefont {Zhang}}, \bibinfo
  {author} {\bibfnamefont {H.}~\bibnamefont {Luo}}, \bibinfo {author}
  {\bibfnamefont {A.~H.}\ \bibnamefont {Nevidomskyy}}, \bibinfo {author}
  {\bibfnamefont {Q.}~\bibnamefont {Si}}, \ and\ \bibinfo {author}
  {\bibfnamefont {P.}~\bibnamefont {Dai}},\ }\href {\doibase
  10.1126/science.1251853} {\bibfield  {journal} {\bibinfo  {journal}
  {Science}\ }\textbf {\bibinfo {volume} {345}},\ \bibinfo {pages} {657 }
  (\bibinfo {year} {2014})}\BibitemShut {NoStop}%
\bibitem [{\citenamefont {Zhang}\ \emph {et~al.}(2015)\citenamefont {Zhang},
  \citenamefont {Fernandes}, \citenamefont {Lamsal}, \citenamefont {Yan},
  \citenamefont {Chi}, \citenamefont {Tucker}, \citenamefont {Pratt},
  \citenamefont {Lynn}, \citenamefont {McCallum}, \citenamefont {Canfield},
  \citenamefont {Lograsso}, \citenamefont {Goldman}, \citenamefont {Vaknin},\
  and\ \citenamefont {McQueeney}}]{GoldmanSpinNematicOrderPRL2015}%
  \BibitemOpen
  \bibfield  {author} {\bibinfo {author} {\bibfnamefont {Q.}~\bibnamefont
  {Zhang}}, \bibinfo {author} {\bibfnamefont {R.~M.}\ \bibnamefont
  {Fernandes}}, \bibinfo {author} {\bibfnamefont {J.}~\bibnamefont {Lamsal}},
  \bibinfo {author} {\bibfnamefont {J.}~\bibnamefont {Yan}}, \bibinfo {author}
  {\bibfnamefont {S.}~\bibnamefont {Chi}}, \bibinfo {author} {\bibfnamefont
  {G.~S.}\ \bibnamefont {Tucker}}, \bibinfo {author} {\bibfnamefont {D.~K.}\
  \bibnamefont {Pratt}}, \bibinfo {author} {\bibfnamefont {J.~W.}\ \bibnamefont
  {Lynn}}, \bibinfo {author} {\bibfnamefont {R.~W.}\ \bibnamefont {McCallum}},
  \bibinfo {author} {\bibfnamefont {P.~C.}\ \bibnamefont {Canfield}}, \bibinfo
  {author} {\bibfnamefont {T.~A.}\ \bibnamefont {Lograsso}}, \bibinfo {author}
  {\bibfnamefont {A.~I.}\ \bibnamefont {Goldman}}, \bibinfo {author}
  {\bibfnamefont {D.}~\bibnamefont {Vaknin}}, \ and\ \bibinfo {author}
  {\bibfnamefont {R.~J.}\ \bibnamefont {McQueeney}},\ }\href {\doibase
  10.1103/PhysRevLett.114.057001} {\bibfield  {journal} {\bibinfo  {journal}
  {Phys. Rev. Lett.}\ }\textbf {\bibinfo {volume} {114}},\ \bibinfo {pages}
  {057001} (\bibinfo {year} {2015})}\BibitemShut {NoStop}%
\bibitem [{\citenamefont {Baek}\ \emph {et~al.}(2015)\citenamefont {Baek},
  \citenamefont {Efremov}, \citenamefont {Ok}, \citenamefont {Kim},
  \citenamefont {van~den Brink},\ and\ \citenamefont {B\"{u}chner}}]{Baek2015}%
  \BibitemOpen
  \bibfield  {author} {\bibinfo {author} {\bibfnamefont {S.-H.}\ \bibnamefont
  {Baek}}, \bibinfo {author} {\bibfnamefont {D.~V.}\ \bibnamefont {Efremov}},
  \bibinfo {author} {\bibfnamefont {J.~M.}\ \bibnamefont {Ok}}, \bibinfo
  {author} {\bibfnamefont {J.~S.}\ \bibnamefont {Kim}}, \bibinfo {author}
  {\bibfnamefont {J.}~\bibnamefont {van~den Brink}}, \ and\ \bibinfo {author}
  {\bibfnamefont {B.}~\bibnamefont {B\"{u}chner}},\ }\href
  {http://dx.doi.org/10.1038/nmat4138} {\bibfield  {journal} {\bibinfo
  {journal} {Nat. Mater.}\ }\textbf {\bibinfo {volume} {14}},\ \bibinfo {pages}
  {210} (\bibinfo {year} {2015})}\BibitemShut {NoStop}%
\bibitem [{\citenamefont {B\"ohmer}\ \emph {et~al.}(2015)\citenamefont
  {B\"ohmer}, \citenamefont {Arai}, \citenamefont {Hardy}, \citenamefont
  {Hattori}, \citenamefont {Iye}, \citenamefont {Wolf}, \citenamefont
  {L\"ohneysen}, \citenamefont {Ishida},\ and\ \citenamefont
  {Meingast}}]{IshidaFeSePRL2015}%
  \BibitemOpen
  \bibfield  {author} {\bibinfo {author} {\bibfnamefont {A.~E.}\ \bibnamefont
  {B\"ohmer}}, \bibinfo {author} {\bibfnamefont {T.}~\bibnamefont {Arai}},
  \bibinfo {author} {\bibfnamefont {F.}~\bibnamefont {Hardy}}, \bibinfo
  {author} {\bibfnamefont {T.}~\bibnamefont {Hattori}}, \bibinfo {author}
  {\bibfnamefont {T.}~\bibnamefont {Iye}}, \bibinfo {author} {\bibfnamefont
  {T.}~\bibnamefont {Wolf}}, \bibinfo {author} {\bibfnamefont {H.~v.}\
  \bibnamefont {L\"ohneysen}}, \bibinfo {author} {\bibfnamefont
  {K.}~\bibnamefont {Ishida}}, \ and\ \bibinfo {author} {\bibfnamefont
  {C.}~\bibnamefont {Meingast}},\ }\href {\doibase
  10.1103/PhysRevLett.114.027001} {\bibfield  {journal} {\bibinfo  {journal}
  {Phys. Rev. Lett.}\ }\textbf {\bibinfo {volume} {114}},\ \bibinfo {pages}
  {027001} (\bibinfo {year} {2015})}\BibitemShut {NoStop}%
\bibitem [{\citenamefont {Rahn}\ \emph {et~al.}(2015)\citenamefont {Rahn},
  \citenamefont {Ewings}, \citenamefont {Sedlmaier}, \citenamefont {Clarke},\
  and\ \citenamefont {Boothroyd}}]{Boothroyd_FeSe}%
  \BibitemOpen
  \bibfield  {author} {\bibinfo {author} {\bibfnamefont {M.~C.}\ \bibnamefont
  {Rahn}}, \bibinfo {author} {\bibfnamefont {R.~A.}\ \bibnamefont {Ewings}},
  \bibinfo {author} {\bibfnamefont {S.~J.}\ \bibnamefont {Sedlmaier}}, \bibinfo
  {author} {\bibfnamefont {S.~J.}\ \bibnamefont {Clarke}}, \ and\ \bibinfo
  {author} {\bibfnamefont {A.~T.}\ \bibnamefont {Boothroyd}},\ }\href {\doibase
  10.1103/PhysRevB.91.180501} {\bibfield  {journal} {\bibinfo  {journal} {Phys.
  Rev. B}\ }\textbf {\bibinfo {volume} {91}},\ \bibinfo {pages} {180501}
  (\bibinfo {year} {2015})}\BibitemShut {NoStop}%
\bibitem [{\citenamefont {Wang}\ \emph {et~al.}(2015)\citenamefont {Wang},
  \citenamefont {Shen}, \citenamefont {Pan}, \citenamefont {Hao}, \citenamefont
  {Ma}, \citenamefont {Zhou}, \citenamefont {Steffens}, \citenamefont
  {Schmalzl}, \citenamefont {Forrest}, \citenamefont {Abdel-Hafiez} \emph
  {et~al.}}]{Wang_FeSe}%
  \BibitemOpen
  \bibfield  {author} {\bibinfo {author} {\bibfnamefont {Q.}~\bibnamefont
  {Wang}}, \bibinfo {author} {\bibfnamefont {Y.}~\bibnamefont {Shen}}, \bibinfo
  {author} {\bibfnamefont {B.}~\bibnamefont {Pan}}, \bibinfo {author}
  {\bibfnamefont {Y.}~\bibnamefont {Hao}}, \bibinfo {author} {\bibfnamefont
  {M.}~\bibnamefont {Ma}}, \bibinfo {author} {\bibfnamefont {F.}~\bibnamefont
  {Zhou}}, \bibinfo {author} {\bibfnamefont {P.}~\bibnamefont {Steffens}},
  \bibinfo {author} {\bibfnamefont {K.}~\bibnamefont {Schmalzl}}, \bibinfo
  {author} {\bibfnamefont {T.}~\bibnamefont {Forrest}}, \bibinfo {author}
  {\bibfnamefont {M.}~\bibnamefont {Abdel-Hafiez}},  \emph {et~al.},\
  }\href@noop {} {\bibfield  {journal} {\bibinfo  {journal} {arXiv preprint
  arXiv:1502.07544}\ } (\bibinfo {year} {2015})}\BibitemShut {NoStop}%
\bibitem [{\citenamefont {Dioguardi}\ \emph {et~al.}(2013)\citenamefont
  {Dioguardi}, \citenamefont {Crocker}, \citenamefont {Shockley}, \citenamefont
  {Lin}, \citenamefont {Shirer}, \citenamefont {Nisson}, \citenamefont
  {Lawson}, \citenamefont {apRoberts Warren}, \citenamefont {Canfield},
  \citenamefont {Bud'ko}, \citenamefont {ran},\ and\ \citenamefont
  {Curro}}]{Dioguardi:2013dd}%
  \BibitemOpen
  \bibfield  {author} {\bibinfo {author} {\bibfnamefont {A.~P.}\ \bibnamefont
  {Dioguardi}}, \bibinfo {author} {\bibfnamefont {J.}~\bibnamefont {Crocker}},
  \bibinfo {author} {\bibfnamefont {A.~C.}\ \bibnamefont {Shockley}}, \bibinfo
  {author} {\bibfnamefont {C.~H.}\ \bibnamefont {Lin}}, \bibinfo {author}
  {\bibfnamefont {K.~R.}\ \bibnamefont {Shirer}}, \bibinfo {author}
  {\bibfnamefont {D.~M.}\ \bibnamefont {Nisson}}, \bibinfo {author}
  {\bibfnamefont {M.~M.}\ \bibnamefont {Lawson}}, \bibinfo {author}
  {\bibfnamefont {N.}~\bibnamefont {apRoberts Warren}}, \bibinfo {author}
  {\bibfnamefont {P.~C.}\ \bibnamefont {Canfield}}, \bibinfo {author}
  {\bibfnamefont {S.~L.}\ \bibnamefont {Bud'ko}}, \bibinfo {author}
  {\bibfnamefont {S.}~\bibnamefont {ran}}, \ and\ \bibinfo {author}
  {\bibfnamefont {N.~J.}\ \bibnamefont {Curro}},\ }\href {\doibase
  10.1103/PhysRevLett.111.207201} {\bibfield  {journal} {\bibinfo  {journal}
  {Phys. Rev. Lett.}\ }\textbf {\bibinfo {volume} {111}},\ \bibinfo {pages}
  {207201} (\bibinfo {year} {2013})}\BibitemShut {NoStop}%
\bibitem [{\citenamefont {Dioguardi}\ \emph {et~al.}(2015)\citenamefont
  {Dioguardi}, \citenamefont {Lawson}, \citenamefont {Bush}, \citenamefont
  {Crocker}, \citenamefont {Shirer}, \citenamefont {Nisson}, \citenamefont
  {Kissikov}, \citenamefont {ran}, \citenamefont {Bud'ko}, \citenamefont
  {Canfield}, \citenamefont {Yuan}, \citenamefont {Kuhns}, \citenamefont
  {Reyes}, \citenamefont {Grafe},\ and\ \citenamefont
  {Curro}}]{Dioguardi:2015tx}%
  \BibitemOpen
  \bibfield  {author} {\bibinfo {author} {\bibfnamefont {A.~P.}\ \bibnamefont
  {Dioguardi}}, \bibinfo {author} {\bibfnamefont {M.~M.}\ \bibnamefont
  {Lawson}}, \bibinfo {author} {\bibfnamefont {B.~T.}\ \bibnamefont {Bush}},
  \bibinfo {author} {\bibfnamefont {J.}~\bibnamefont {Crocker}}, \bibinfo
  {author} {\bibfnamefont {K.~R.}\ \bibnamefont {Shirer}}, \bibinfo {author}
  {\bibfnamefont {D.~M.}\ \bibnamefont {Nisson}}, \bibinfo {author}
  {\bibfnamefont {T.}~\bibnamefont {Kissikov}}, \bibinfo {author}
  {\bibfnamefont {S.}~\bibnamefont {ran}}, \bibinfo {author} {\bibfnamefont
  {S.~L.}\ \bibnamefont {Bud'ko}}, \bibinfo {author} {\bibfnamefont {P.~C.}\
  \bibnamefont {Canfield}}, \bibinfo {author} {\bibfnamefont {S.}~\bibnamefont
  {Yuan}}, \bibinfo {author} {\bibfnamefont {P.~L.}\ \bibnamefont {Kuhns}},
  \bibinfo {author} {\bibfnamefont {A.~P.}\ \bibnamefont {Reyes}}, \bibinfo
  {author} {\bibfnamefont {H.~J.}\ \bibnamefont {Grafe}}, \ and\ \bibinfo
  {author} {\bibfnamefont {N.~J.}\ \bibnamefont {Curro}},\ }\href
  {http://arxiv.org/abs/1503.01844} {\bibfield  {journal} {\bibinfo  {journal}
  {arXiv.org}\ } (\bibinfo {year} {2015})},\ \Eprint
  {http://arxiv.org/abs/1503.01844} {1503.01844} \BibitemShut {NoStop}%
\bibitem [{\citenamefont {Loh}\ \emph {et~al.}(2010)\citenamefont {Loh},
  \citenamefont {Carlson},\ and\ \citenamefont {Dahmen}}]{RFIMsimulations2010}%
  \BibitemOpen
  \bibfield  {author} {\bibinfo {author} {\bibnamefont {Loh}}, \bibinfo
  {author} {\bibfnamefont {E.~W.}\ \bibnamefont {Carlson}}, \ and\ \bibinfo
  {author} {\bibfnamefont {K.~A.}\ \bibnamefont {Dahmen}},\ }\href {\doibase
  10.1103/PhysRevB.81.224207} {\bibfield  {journal} {\bibinfo  {journal}
  {Phys.Rev. B}\ }\textbf {\bibinfo {volume} {81}},\ \bibinfo {pages} {224207}
  (\bibinfo {year} {2010})}\BibitemShut {NoStop}%
\bibitem [{\citenamefont {Nakai}\ \emph {et~al.}(2010)\citenamefont {Nakai},
  \citenamefont {Nakai}, \citenamefont {Iye}, \citenamefont {Kitagawa},
  \citenamefont {Iye}, \citenamefont {Ishida}, \citenamefont {Kitagawa},
  \citenamefont {Ikeda}, \citenamefont {Ishida}, \citenamefont {Kasahara},
  \citenamefont {Ikeda}, \citenamefont {Shishido}, \citenamefont {Shishido},
  \citenamefont {Shibauchi}, \citenamefont {Matsuda},\ and\ \citenamefont
  {Terashima}}]{Nakai:2010ku}%
  \BibitemOpen
  \bibfield  {author} {\bibinfo {author} {\bibfnamefont {Y.}~\bibnamefont
  {Nakai}}, \bibinfo {author} {\bibfnamefont {Y.}~\bibnamefont {Nakai}},
  \bibinfo {author} {\bibfnamefont {T.}~\bibnamefont {Iye}}, \bibinfo {author}
  {\bibfnamefont {S.}~\bibnamefont {Kitagawa}}, \bibinfo {author}
  {\bibfnamefont {T.}~\bibnamefont {Iye}}, \bibinfo {author} {\bibfnamefont
  {K.}~\bibnamefont {Ishida}}, \bibinfo {author} {\bibfnamefont
  {S.}~\bibnamefont {Kitagawa}}, \bibinfo {author} {\bibfnamefont
  {H.}~\bibnamefont {Ikeda}}, \bibinfo {author} {\bibfnamefont
  {K.}~\bibnamefont {Ishida}}, \bibinfo {author} {\bibfnamefont
  {S.}~\bibnamefont {Kasahara}}, \bibinfo {author} {\bibfnamefont
  {H.}~\bibnamefont {Ikeda}}, \bibinfo {author} {\bibfnamefont
  {H.}~\bibnamefont {Shishido}}, \bibinfo {author} {\bibfnamefont
  {H.}~\bibnamefont {Shishido}}, \bibinfo {author} {\bibfnamefont
  {T.}~\bibnamefont {Shibauchi}}, \bibinfo {author} {\bibfnamefont
  {Y.}~\bibnamefont {Matsuda}}, \ and\ \bibinfo {author} {\bibfnamefont
  {T.}~\bibnamefont {Terashima}},\ }\href {\doibase
  10.1103/physrevlett.105.107003} {\bibfield  {journal} {\bibinfo  {journal}
  {Phys. Rev. Lett.}\ }\textbf {\bibinfo {volume} {105}},\ \bibinfo {pages}
  {107003} (\bibinfo {year} {2010})}\BibitemShut {NoStop}%
\bibitem [{\citenamefont {Nakai}\ \emph {et~al.}(2013)\citenamefont {Nakai},
  \citenamefont {Iye}, \citenamefont {Kitagawa}, \citenamefont {Iye},
  \citenamefont {Ishida}, \citenamefont {Kitagawa}, \citenamefont {Kasahara},
  \citenamefont {Shibauchi}, \citenamefont {Matsuda}, \citenamefont {Ikeda},\
  and\ \citenamefont {Terashima}}]{Nakai:2013by}%
  \BibitemOpen
  \bibfield  {author} {\bibinfo {author} {\bibfnamefont {Y.}~\bibnamefont
  {Nakai}}, \bibinfo {author} {\bibfnamefont {T.}~\bibnamefont {Iye}}, \bibinfo
  {author} {\bibfnamefont {S.}~\bibnamefont {Kitagawa}}, \bibinfo {author}
  {\bibfnamefont {T.}~\bibnamefont {Iye}}, \bibinfo {author} {\bibfnamefont
  {K.}~\bibnamefont {Ishida}}, \bibinfo {author} {\bibfnamefont
  {S.}~\bibnamefont {Kitagawa}}, \bibinfo {author} {\bibfnamefont
  {S.}~\bibnamefont {Kasahara}}, \bibinfo {author} {\bibfnamefont
  {T.}~\bibnamefont {Shibauchi}}, \bibinfo {author} {\bibfnamefont
  {Y.}~\bibnamefont {Matsuda}}, \bibinfo {author} {\bibfnamefont
  {H.}~\bibnamefont {Ikeda}}, \ and\ \bibinfo {author} {\bibfnamefont
  {T.}~\bibnamefont {Terashima}},\ }\href {\doibase 10.1103/physrevb.87.174507}
  {\bibfield  {journal} {\bibinfo  {journal} {Phys. Rev. B}\ }\textbf {\bibinfo
  {volume} {87}},\ \bibinfo {pages} {174507} (\bibinfo {year}
  {2013})}\BibitemShut {NoStop}%
\bibitem [{\citenamefont {Hammerath}\ \emph {et~al.}(2010)\citenamefont
  {Hammerath}, \citenamefont {Drechsler}, \citenamefont {Grafe}, \citenamefont
  {Lang}, \citenamefont {Fuchs}, \citenamefont {Behr}, \citenamefont {Eremin},
  \citenamefont {Korshunov},\ and\ \citenamefont
  {B{\"u}chner}}]{hammerath:2010ki}%
  \BibitemOpen
  \bibfield  {author} {\bibinfo {author} {\bibfnamefont {F.}~\bibnamefont
  {Hammerath}}, \bibinfo {author} {\bibfnamefont {S.~L.}\ \bibnamefont
  {Drechsler}}, \bibinfo {author} {\bibfnamefont {H.~J.}\ \bibnamefont
  {Grafe}}, \bibinfo {author} {\bibfnamefont {G.}~\bibnamefont {Lang}},
  \bibinfo {author} {\bibfnamefont {G.}~\bibnamefont {Fuchs}}, \bibinfo
  {author} {\bibfnamefont {G.}~\bibnamefont {Behr}}, \bibinfo {author}
  {\bibfnamefont {I.}~\bibnamefont {Eremin}}, \bibinfo {author} {\bibfnamefont
  {M.~M.}\ \bibnamefont {Korshunov}}, \ and\ \bibinfo {author} {\bibfnamefont
  {B.}~\bibnamefont {B{\"u}chner}},\ }\href {\doibase
  10.1103/PhysRevB.81.140504} {\bibfield  {journal} {\bibinfo  {journal} {Phys.
  Rev. B}\ }\textbf {\bibinfo {volume} {81}},\ \bibinfo {pages} {140504}
  (\bibinfo {year} {2010})}\BibitemShut {NoStop}%
\bibitem [{\citenamefont {Hammerath}\ \emph {et~al.}(2013)\citenamefont
  {Hammerath}, \citenamefont {Gr{\"a}fe}, \citenamefont {K{\"u}hne},
  \citenamefont {K{\"u}hne}, \citenamefont {Kuhns}, \citenamefont {Reyes},
  \citenamefont {Lang}, \citenamefont {Wurmehl}, \citenamefont {B{\"u}chner},
  \citenamefont {Carretta},\ and\ \citenamefont {Grafe}}]{hammerath:2013jj}%
  \BibitemOpen
  \bibfield  {author} {\bibinfo {author} {\bibfnamefont {F.}~\bibnamefont
  {Hammerath}}, \bibinfo {author} {\bibfnamefont {U.}~\bibnamefont
  {Gr{\"a}fe}}, \bibinfo {author} {\bibfnamefont {T.}~\bibnamefont
  {K{\"u}hne}}, \bibinfo {author} {\bibfnamefont {H.}~\bibnamefont
  {K{\"u}hne}}, \bibinfo {author} {\bibfnamefont {P.~L.}\ \bibnamefont
  {Kuhns}}, \bibinfo {author} {\bibfnamefont {A.~P.}\ \bibnamefont {Reyes}},
  \bibinfo {author} {\bibfnamefont {G.}~\bibnamefont {Lang}}, \bibinfo {author}
  {\bibfnamefont {S.}~\bibnamefont {Wurmehl}}, \bibinfo {author} {\bibfnamefont
  {B.}~\bibnamefont {B{\"u}chner}}, \bibinfo {author} {\bibfnamefont
  {P.}~\bibnamefont {Carretta}}, \ and\ \bibinfo {author} {\bibfnamefont
  {H.~J.}\ \bibnamefont {Grafe}},\ }\href {\doibase 10.1103/PhysRevB.88.104503}
  {\bibfield  {journal} {\bibinfo  {journal} {Phys. Rev. B}\ }\textbf {\bibinfo
  {volume} {88}},\ \bibinfo {pages} {104503} (\bibinfo {year}
  {2013})}\BibitemShut {NoStop}%
\bibitem [{\citenamefont {Kuo}\ \emph {et~al.}(2012)\citenamefont {Kuo},
  \citenamefont {Analytis}, \citenamefont {Chu}, \citenamefont {Fernandes},
  \citenamefont {Schmalian},\ and\ \citenamefont {Fisher}}]{KuoPBa122PRB2012}%
  \BibitemOpen
  \bibfield  {author} {\bibinfo {author} {\bibfnamefont {H.-H.}\ \bibnamefont
  {Kuo}}, \bibinfo {author} {\bibfnamefont {J.~G.}\ \bibnamefont {Analytis}},
  \bibinfo {author} {\bibfnamefont {J.-H.}\ \bibnamefont {Chu}}, \bibinfo
  {author} {\bibfnamefont {R.~M.}\ \bibnamefont {Fernandes}}, \bibinfo {author}
  {\bibfnamefont {J.}~\bibnamefont {Schmalian}}, \ and\ \bibinfo {author}
  {\bibfnamefont {I.~R.}\ \bibnamefont {Fisher}},\ }\href {\doibase
  10.1103/PhysRevB.86.134507} {\bibfield  {journal} {\bibinfo  {journal} {Phys.
  Rev. B}\ }\textbf {\bibinfo {volume} {86}},\ \bibinfo {pages} {134507}
  (\bibinfo {year} {2012})}\BibitemShut {NoStop}%
\bibitem [{\citenamefont {Johnston}(2006)}]{Johnston:2006gs}%
  \BibitemOpen
  \bibfield  {author} {\bibinfo {author} {\bibfnamefont {D.}~\bibnamefont
  {Johnston}},\ }\href {\doibase 10.1103/PhysRevB.74.184430} {\bibfield
  {journal} {\bibinfo  {journal} {Phys. Rev. B}\ }\textbf {\bibinfo {volume}
  {74}},\ \bibinfo {pages} {184430} (\bibinfo {year} {2006})}\BibitemShut
  {NoStop}%
\bibitem [{\citenamefont {Kitagawa}\ \emph {et~al.}(2008)\citenamefont
  {Kitagawa}, \citenamefont {Katayama}, \citenamefont {Ohgushi}, \citenamefont
  {Yoshida},\ and\ \citenamefont {Takigawa}}]{takigawa2008}%
  \BibitemOpen
  \bibfield  {author} {\bibinfo {author} {\bibfnamefont {K.}~\bibnamefont
  {Kitagawa}}, \bibinfo {author} {\bibfnamefont {N.}~\bibnamefont {Katayama}},
  \bibinfo {author} {\bibfnamefont {K.}~\bibnamefont {Ohgushi}}, \bibinfo
  {author} {\bibfnamefont {M.}~\bibnamefont {Yoshida}}, \ and\ \bibinfo
  {author} {\bibfnamefont {M.}~\bibnamefont {Takigawa}},\ }\href {\doibase
  10.1143/JPSJ.77.114709} {\bibfield  {journal} {\bibinfo  {journal} {J. Phys.
  Soc. Jpn.}\ }\textbf {\bibinfo {volume} {77}},\ \bibinfo {pages} {114709}
  (\bibinfo {year} {2008})}\BibitemShut {NoStop}%
\bibitem [{\citenamefont {Slichter}(1990)}]{slichter1990principles}%
  \BibitemOpen
  \bibfield  {author} {\bibinfo {author} {\bibfnamefont {C.~P.}\ \bibnamefont
  {Slichter}},\ }\href {http://books.google.com/books?id=fQDvAAAAMAAJ} {\emph
  {\bibinfo {title} {{Principles of magnetic resonance}}}},\ Springer series in
  solid-state sciences\ (\bibinfo  {publisher} {Springer-Verlag},\ \bibinfo
  {year} {1990})\BibitemShut {NoStop}%
\bibitem [{\citenamefont {Kinouchi}\ \emph {et~al.}(2013)\citenamefont
  {Kinouchi}, \citenamefont {Mukuda}, \citenamefont {Kitaoka}, \citenamefont
  {Shirage}, \citenamefont {Fujihisa}, \citenamefont {Gotoh}, \citenamefont
  {Eisaki},\ and\ \citenamefont {Iyo}}]{Kinouchi2012}%
  \BibitemOpen
  \bibfield  {author} {\bibinfo {author} {\bibfnamefont {H.}~\bibnamefont
  {Kinouchi}}, \bibinfo {author} {\bibfnamefont {H.}~\bibnamefont {Mukuda}},
  \bibinfo {author} {\bibfnamefont {Y.}~\bibnamefont {Kitaoka}}, \bibinfo
  {author} {\bibfnamefont {P.~M.}\ \bibnamefont {Shirage}}, \bibinfo {author}
  {\bibfnamefont {H.}~\bibnamefont {Fujihisa}}, \bibinfo {author}
  {\bibfnamefont {Y.}~\bibnamefont {Gotoh}}, \bibinfo {author} {\bibfnamefont
  {H.}~\bibnamefont {Eisaki}}, \ and\ \bibinfo {author} {\bibfnamefont
  {A.}~\bibnamefont {Iyo}},\ }\href {\doibase 10.1103/PhysRevB.87.121101}
  {\bibfield  {journal} {\bibinfo  {journal} {Phys. Rev. B}\ }\textbf {\bibinfo
  {volume} {87}},\ \bibinfo {pages} {121101} (\bibinfo {year}
  {2013})}\BibitemShut {NoStop}%
\bibitem [{\citenamefont {Smerald}\ and\ \citenamefont
  {Shannon}(2011)}]{T1formfactorsArsenides}%
  \BibitemOpen
  \bibfield  {author} {\bibinfo {author} {\bibfnamefont {A.}~\bibnamefont
  {Smerald}}\ and\ \bibinfo {author} {\bibfnamefont {N.}~\bibnamefont
  {Shannon}},\ }\href {\doibase 10.1103/PhysRevB.84.184437} {\bibfield
  {journal} {\bibinfo  {journal} {Phys. Rev. B}\ }\textbf {\bibinfo {volume}
  {84}},\ \bibinfo {pages} {184437} (\bibinfo {year} {2011})}\BibitemShut
  {NoStop}%
\bibitem [{\citenamefont {Millis}\ \emph {et~al.}(1990)\citenamefont {Millis},
  \citenamefont {Monien},\ and\ \citenamefont {Pines}}]{MMPT1inYBCO}%
  \BibitemOpen
  \bibfield  {author} {\bibinfo {author} {\bibfnamefont {A.~J.}\ \bibnamefont
  {Millis}}, \bibinfo {author} {\bibfnamefont {H.}~\bibnamefont {Monien}}, \
  and\ \bibinfo {author} {\bibfnamefont {D.}~\bibnamefont {Pines}},\ }\href
  {\doibase 10.1103/PhysRevB.42.167} {\bibfield  {journal} {\bibinfo  {journal}
  {Phys. Rev. B}\ }\textbf {\bibinfo {volume} {42}},\ \bibinfo {pages} {167}
  (\bibinfo {year} {1990})}\BibitemShut {NoStop}%
\bibitem [{\citenamefont {Suter}\ \emph {et~al.}(1999)\citenamefont {Suter},
  \citenamefont {Mali}, \citenamefont {Roos},\ and\ \citenamefont
  {Brinkmann}}]{Suter:1999kw}%
  \BibitemOpen
  \bibfield  {author} {\bibinfo {author} {\bibfnamefont {A.}~\bibnamefont
  {Suter}}, \bibinfo {author} {\bibfnamefont {M.}~\bibnamefont {Mali}},
  \bibinfo {author} {\bibfnamefont {J.}~\bibnamefont {Roos}}, \ and\ \bibinfo
  {author} {\bibfnamefont {D.}~\bibnamefont {Brinkmann}},\ }\href {\doibase
  10.1088/0953-8984/10/26/022} {\bibfield  {journal} {\bibinfo  {journal}
  {Supercond. Sci. Technol.}\ }\textbf {\bibinfo {volume} {10}},\ \bibinfo
  {pages} {5977} (\bibinfo {year} {1999})}\BibitemShut {NoStop}%
\bibitem [{\citenamefont {Iye}\ \emph {et~al.}(2012)\citenamefont {Iye},
  \citenamefont {Nakai}, \citenamefont {Kitagawa}, \citenamefont {Ishida},
  \citenamefont {Kasahara}, \citenamefont {Shibauchi}, \citenamefont
  {Matsuda},\ and\ \citenamefont {Terashima}}]{IshidaPdopedBa122JPSJ2012}%
  \BibitemOpen
  \bibfield  {author} {\bibinfo {author} {\bibfnamefont {T.}~\bibnamefont
  {Iye}}, \bibinfo {author} {\bibfnamefont {Y.}~\bibnamefont {Nakai}}, \bibinfo
  {author} {\bibfnamefont {S.}~\bibnamefont {Kitagawa}}, \bibinfo {author}
  {\bibfnamefont {K.}~\bibnamefont {Ishida}}, \bibinfo {author} {\bibfnamefont
  {S.}~\bibnamefont {Kasahara}}, \bibinfo {author} {\bibfnamefont
  {T.}~\bibnamefont {Shibauchi}}, \bibinfo {author} {\bibfnamefont
  {Y.}~\bibnamefont {Matsuda}}, \ and\ \bibinfo {author} {\bibfnamefont
  {T.}~\bibnamefont {Terashima}},\ }\href {\doibase 10.1143/JPSJ.81.033701}
  {\bibfield  {journal} {\bibinfo  {journal} {J. Phys. Soc. Jpn.}\ }\textbf
  {\bibinfo {volume} {81}},\ \bibinfo {pages} {033701} (\bibinfo {year}
  {2012})}\BibitemShut {NoStop}%
\bibitem [{\citenamefont {Hu}\ \emph {et~al.}(2015)\citenamefont {Hu},
  \citenamefont {Lu}, \citenamefont {Zhang}, \citenamefont {Luo}, \citenamefont
  {Li}, \citenamefont {Wang}, \citenamefont {Chen}, \citenamefont {Han},
  \citenamefont {Banjara}, \citenamefont {Sapkota}, \citenamefont {Kreyssig},
  \citenamefont {Goldman}, \citenamefont {Yamani}, \citenamefont {Niedermayer},
  \citenamefont {Skoulatos}, \citenamefont {Georgii}, \citenamefont {Keller},
  \citenamefont {Wang}, \citenamefont {Yu},\ and\ \citenamefont
  {Dai}}]{PBa122StructureMagneticPRL2015}%
  \BibitemOpen
  \bibfield  {author} {\bibinfo {author} {\bibfnamefont {D.}~\bibnamefont
  {Hu}}, \bibinfo {author} {\bibfnamefont {X.}~\bibnamefont {Lu}}, \bibinfo
  {author} {\bibfnamefont {W.}~\bibnamefont {Zhang}}, \bibinfo {author}
  {\bibfnamefont {H.}~\bibnamefont {Luo}}, \bibinfo {author} {\bibfnamefont
  {S.}~\bibnamefont {Li}}, \bibinfo {author} {\bibfnamefont {P.}~\bibnamefont
  {Wang}}, \bibinfo {author} {\bibfnamefont {G.}~\bibnamefont {Chen}}, \bibinfo
  {author} {\bibfnamefont {F.}~\bibnamefont {Han}}, \bibinfo {author}
  {\bibfnamefont {S.~R.}\ \bibnamefont {Banjara}}, \bibinfo {author}
  {\bibfnamefont {A.}~\bibnamefont {Sapkota}}, \bibinfo {author} {\bibfnamefont
  {A.}~\bibnamefont {Kreyssig}}, \bibinfo {author} {\bibfnamefont {A.~I.}\
  \bibnamefont {Goldman}}, \bibinfo {author} {\bibfnamefont {Z.}~\bibnamefont
  {Yamani}}, \bibinfo {author} {\bibfnamefont {C.}~\bibnamefont {Niedermayer}},
  \bibinfo {author} {\bibfnamefont {M.}~\bibnamefont {Skoulatos}}, \bibinfo
  {author} {\bibfnamefont {R.}~\bibnamefont {Georgii}}, \bibinfo {author}
  {\bibfnamefont {T.}~\bibnamefont {Keller}}, \bibinfo {author} {\bibfnamefont
  {P.}~\bibnamefont {Wang}}, \bibinfo {author} {\bibfnamefont {W.}~\bibnamefont
  {Yu}}, \ and\ \bibinfo {author} {\bibfnamefont {P.}~\bibnamefont {Dai}},\
  }\href {\doibase 10.1103/PhysRevLett.114.157002} {\bibfield  {journal}
  {\bibinfo  {journal} {Phys. Rev. Lett.}\ }\textbf {\bibinfo {volume} {114}},\
  \bibinfo {pages} {157002} (\bibinfo {year} {2015})}\BibitemShut {NoStop}%
\bibitem [{\citenamefont {Takeda}\ \emph {et~al.}(2014)\citenamefont {Takeda},
  \citenamefont {Imai}, \citenamefont {Tachibana}, \citenamefont {Gaudet},
  \citenamefont {Gaulin}, \citenamefont {Saparov},\ and\ \citenamefont
  {Sefat}}]{Takeda:2014ia}%
  \BibitemOpen
  \bibfield  {author} {\bibinfo {author} {\bibfnamefont {H.}~\bibnamefont
  {Takeda}}, \bibinfo {author} {\bibfnamefont {T.}~\bibnamefont {Imai}},
  \bibinfo {author} {\bibfnamefont {M.}~\bibnamefont {Tachibana}}, \bibinfo
  {author} {\bibfnamefont {J.}~\bibnamefont {Gaudet}}, \bibinfo {author}
  {\bibfnamefont {B.~D.}\ \bibnamefont {Gaulin}}, \bibinfo {author}
  {\bibfnamefont {B.~I.}\ \bibnamefont {Saparov}}, \ and\ \bibinfo {author}
  {\bibfnamefont {A.~S.}\ \bibnamefont {Sefat}},\ }\href {\doibase
  10.1103/PhysRevLett.113.117001} {\bibfield  {journal} {\bibinfo  {journal}
  {Phys. Rev. Lett.}\ }\textbf {\bibinfo {volume} {113}},\ \bibinfo {pages}
  {117001} (\bibinfo {year} {2014})}\BibitemShut {NoStop}%
\bibitem [{\citenamefont {Schmalian}\ and\ \citenamefont
  {Wolynes}(2000)}]{Schmalian_Wolynes00}%
  \BibitemOpen
  \bibfield  {author} {\bibinfo {author} {\bibfnamefont {J.}~\bibnamefont
  {Schmalian}}\ and\ \bibinfo {author} {\bibfnamefont {P.~G.}\ \bibnamefont
  {Wolynes}},\ }\href {\doibase 10.1103/PhysRevLett.85.836} {\bibfield
  {journal} {\bibinfo  {journal} {Phys. Rev. Lett.}\ }\textbf {\bibinfo
  {volume} {85}},\ \bibinfo {pages} {836} (\bibinfo {year} {2000})}\BibitemShut
  {NoStop}%
\bibitem [{\citenamefont {Kasahara}\ \emph {et~al.}(2012)\citenamefont
  {Kasahara}, \citenamefont {Shi}, \citenamefont {Hashimoto}, \citenamefont
  {Tonegawa}, \citenamefont {Mizukami}, \citenamefont {Shibauchi},
  \citenamefont {Sugimoto}, \citenamefont {Fukuda}, \citenamefont {Terashima},
  \citenamefont {Nevidomskyy},\ and\ \citenamefont {Matsuda}}]{Kasahara2012}%
  \BibitemOpen
  \bibfield  {author} {\bibinfo {author} {\bibfnamefont {S.}~\bibnamefont
  {Kasahara}}, \bibinfo {author} {\bibfnamefont {H.~J.}\ \bibnamefont {Shi}},
  \bibinfo {author} {\bibfnamefont {K.}~\bibnamefont {Hashimoto}}, \bibinfo
  {author} {\bibfnamefont {S.}~\bibnamefont {Tonegawa}}, \bibinfo {author}
  {\bibfnamefont {Y.}~\bibnamefont {Mizukami}}, \bibinfo {author}
  {\bibfnamefont {T.}~\bibnamefont {Shibauchi}}, \bibinfo {author}
  {\bibfnamefont {K.}~\bibnamefont {Sugimoto}}, \bibinfo {author}
  {\bibfnamefont {T.}~\bibnamefont {Fukuda}}, \bibinfo {author} {\bibfnamefont
  {T.}~\bibnamefont {Terashima}}, \bibinfo {author} {\bibfnamefont {A.~H.}\
  \bibnamefont {Nevidomskyy}}, \ and\ \bibinfo {author} {\bibfnamefont
  {Y.}~\bibnamefont {Matsuda}},\ }\href {http://dx.doi.org/10.1038/nature11178}
  {\bibfield  {journal} {\bibinfo  {journal} {Nature}\ }\textbf {\bibinfo
  {volume} {486}},\ \bibinfo {pages} {382} (\bibinfo {year}
  {2012})}\BibitemShut {NoStop}%
\bibitem [{\citenamefont {Thewalt}\ \emph {et~al.}(2015)\citenamefont
  {Thewalt}, \citenamefont {Hinton}, \citenamefont {Hayes}, \citenamefont
  {Helm}, \citenamefont {Lee}, \citenamefont {Analytis},\ and\ \citenamefont
  {Orenstein}}]{Thewalt2015}%
  \BibitemOpen
  \bibfield  {author} {\bibinfo {author} {\bibfnamefont {E.}~\bibnamefont
  {Thewalt}}, \bibinfo {author} {\bibfnamefont {J.~P.}\ \bibnamefont {Hinton}},
  \bibinfo {author} {\bibfnamefont {I.~M.}\ \bibnamefont {Hayes}}, \bibinfo
  {author} {\bibfnamefont {T.}~\bibnamefont {Helm}}, \bibinfo {author}
  {\bibfnamefont {D.~H.}\ \bibnamefont {Lee}}, \bibinfo {author} {\bibfnamefont
  {J.~G.}\ \bibnamefont {Analytis}}, \ and\ \bibinfo {author} {\bibfnamefont
  {J.}~\bibnamefont {Orenstein}},\ }\href@noop {} {\  (\bibinfo {year}
  {2015})},\ \Eprint {http://arxiv.org/abs/1507.03981} {1507.03981}
  \BibitemShut {NoStop}%
\end{thebibliography}
